# A WEB-BASED APPLICATION FOR THE MANAGEMENT OF SEMINAR ASSIGNMENTS

**Cristina Turcu, Cornel Turcu, Evelina Graur**
*The "Stefan cel Mare" University Suceava, Romania*

**ABSTRACT**

*Seminar activities never occur randomly or in a vacuum. Carefully planned and well-organized thematically, they allow both undergraduates and graduates to explore in (more) detail various subject areas of particular interest to every individual or group. Discussion and participation characterize this form of learning, which is often supplemented by in-class presentations of previously assigned tasks. The preparation of assignments often presupposes a lot of self-study and research; it is desirable that the student group enrolled in the same seminar be able to share all points of view, results and conclusions.*

*The present paper presents an information system designed to manage the assignments proposed in a seminar and balance students' individual study load. The system was developed in PHP language with all editable features stored in MySQL database for the best speed and performance results. With a modular organization, the system relies on various entities such as Student, Theme, Keyword, Reference, etc. The implemented procedures are enhanced by a user-friendly graphic interface (this will be described in another paper). The system is easy to use; it eliminates appointment dysfunctions, allows students to have their say in selecting/accepting a task and even to correlate their decision with the proposed date for presentation.*

**Introduction**

The (ever) growing number of students, and hence of individual/group seminar themes or projects, is likely to become a burden on the teaching staff. Under the circumstances, the ratio between the time spent in correcting papers and the number of students has dropped considerably. The solution to this problem resides either in the reduction of the number of students per teacher or in spending more time in face-to-face discussions. Although both methods are realistic, they might incur supplementary staff costs that most universities cannot afford to pay. The student-teacher relationship might deteriorate.

An effective deterrent in this case would be the implementation of new ways of sustaining the communication between the two parties. ICT is the key to the problem.

In spite of an apparently minor disadvantage, namely the elimination of face-to-face interaction, the whole range of applications transforms learning into a round-the-clock activity. Moreover, the distribution or selection of themes may be open up to more negotiation and the records may be accessed any time.

The paper presents an application which is being piloted for the management of seminar assignments in a master's degree program at the "Ştefan cel Mare" University in Suceava.

The first part of the paper approaches several methodological aspects of seminar activities in a master degree's program offered by the Faculty of Letters; the second part is dedicated to the presentation of several implementation solutions.

**Assignments in a master's degree program**

The Faculty of Letters in Suceava currently offers three master's degree programs by coursework. The MA in *Mass Media Semiotics* is a one-year cycle of core courses and seminars. In order to complete the program, the 35 enrolled students will take 12 hours of course and seminar work per week for the first semester, and 13 hours of course and seminar work for the second semester. All activities are scheduled for the weekend because the students enrolled in this program have full-time or part-time jobs. After the coursework in *Computer Mediated Communication* scheduled for the first semester, all students in this program attained an adequate level of computer literacy.

With a total number of 14 hours, the seminar in *News Discourse: Paradigms and Textual Strategies* is scheduled for the second semester. The seminar is designed to provide more insight into issues already covered by previous foundation courses and seminars (e.g. *Introduction to Semiotics, The Discourse of Mass Media, Applied Semiotics).*

A highly selective list of resources was provided for the first meeting, as well as a consistent commentary upon two possible lines to be followed in the study of news: *discourse as process* and *discourse as product*. Hence the teacher's suggestion of two major paradigms coexisting in the understanding and interpretation of news: the media paradigm and the generic paradigm.

With no previous in-depth reading on the subject, the students could not make up their mind when they had to choose among the topics proposed for individual study and class presentation. Some students were not entirely convinced that the teacher's "established paths" where the only ones to follow. They wanted to know whether they could come up with a topic of their own and if their proposed topic would be accepted for presentation. At this level of academic instruction, it is perhaps politically incorrect to assign a theme to a student or a group of students without his or their accord. Moreover, it takes a teacher a lot of inspiration and expertise to come up with as many as 35 topics of comparable complexity.

If the content factor arose much debate, the time factor was regarded as highly discriminatory. The students in charge with the following week's assignment were clearly at a disadvantage. As the seminar is due to end on April 30, 2004, some students have more time to prepare than others. Under the circumstances, it was agreed that all 35 students be granted one week of reading and study before they could decide upon their final choice.

There is a whole range of other intervening factors that should demand our attention: the students' work load per week in the other subjects, the students' capacity to stick to dates and deadlines, the appropriate number of topics for one meeting, the efficiency of feed-back from teacher and fellow students, student-student and teacher-student interaction beyond class.

As the traditional ways of managing seminar activities proved to be unsatisfactory, we have decided to resort to a web-based application that could buy both teachers and students some (more) time to ponder over complex matters, make better decisions and share ideas.

**Implementation Procedure**

The Internet application was developed in PHP language with all editable features stored in MySQL databases.

**The conceptual data model and schema design**

The conceptual data model is of major importance in the design of the database. The conceptual level is the central one, reflecting a specific structuring of data that can be taken over and processed by a database management system. The conceptual schema underlies the conceptual model responsible for defining its properties, the attributes of the entities and their grouping, in keeping with well-established criteria of homogeneity.

Conceptually, database design is reliant on various models. The model *Entity-Relationship* is by far the most popular one, and it needs to be compliant with the following rules:

- the names borne by the entities, correspondences, roles and attributes in the conceptual model and then in the defined database need to be unique
- there cannot be but one association among the same entities

The delineation of the Entity-Relationship model is incumbent on a judicious organization of data, more exactly on observing the normal forms FN1-FN3.

Some of the guiding lines followed in the design of the model are presented below:

- **from the students' perspective (figure 1)**
  - there are two categories of users: administrator(s) and students
  - the access of administrator and students is dependent on authentication: users need to type in their email address and password;
  - a student can view the whole list of available themes and opt for one of them. Some themes may be accompanied by presentation deadlines or dates (for instance, week 6); a planning procedure may be developed for those themes that do not bear such temporal specifications;
  - students are entitled to one or more choices; the limitations are dictated by the administrator;
  - when students are allowed to propose their own themes, they are asked to provide the following information:
    - the title;
    - a summary;
    - keywords;
    - a (selective) list of books and articles used, referred to or recommended for the theme in question
    - the proposed presentation deadline or date (validated by the administrator);
  - the proposed theme cannot be added to the list unless it is accepted by the administrator (teacher). In this case, the administrator ascertains the complexity level and decides upon the maximum number of students sharing the theme;
  - the list of themes may be equally viewed by students and administrator. The list is dynamic in that it can be modified at any time by the introduction or deletion of themes;
  - for each and every theme, students may upload corresponding files; the latter are visible in the file list attached to each theme only if they were approved by the administrator;
  - students may request modifications to their personal data and/or password.

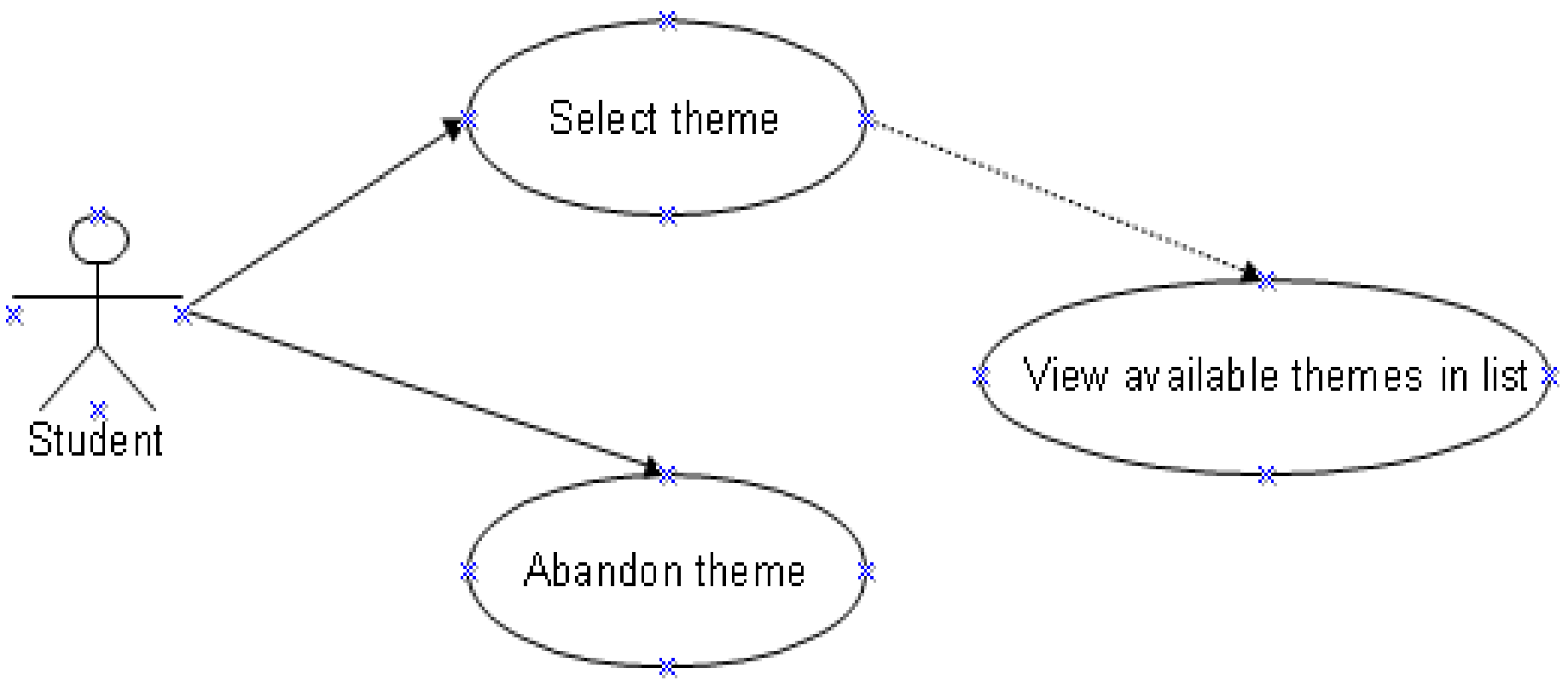

Figure 1. A simplified student case diagram

- **from the administrator's perspective (figure 2)**
  - the administrator can view all students' selections;
  - a list of students sharing the same theme is available for each and every theme;
  - the administrator can view the themes proposed by students;
  - the administrator decides upon the active/inactive status of themes;
  - the administrator may decide to delete a theme in the theme list; if the theme in question is already taken by a student, it cannot be deleted from the database;
  - the administrator is entitled to modify student identification data (name, surname, group, email address); when the password is modified, the user will automatically receive a notification/confirmation email;

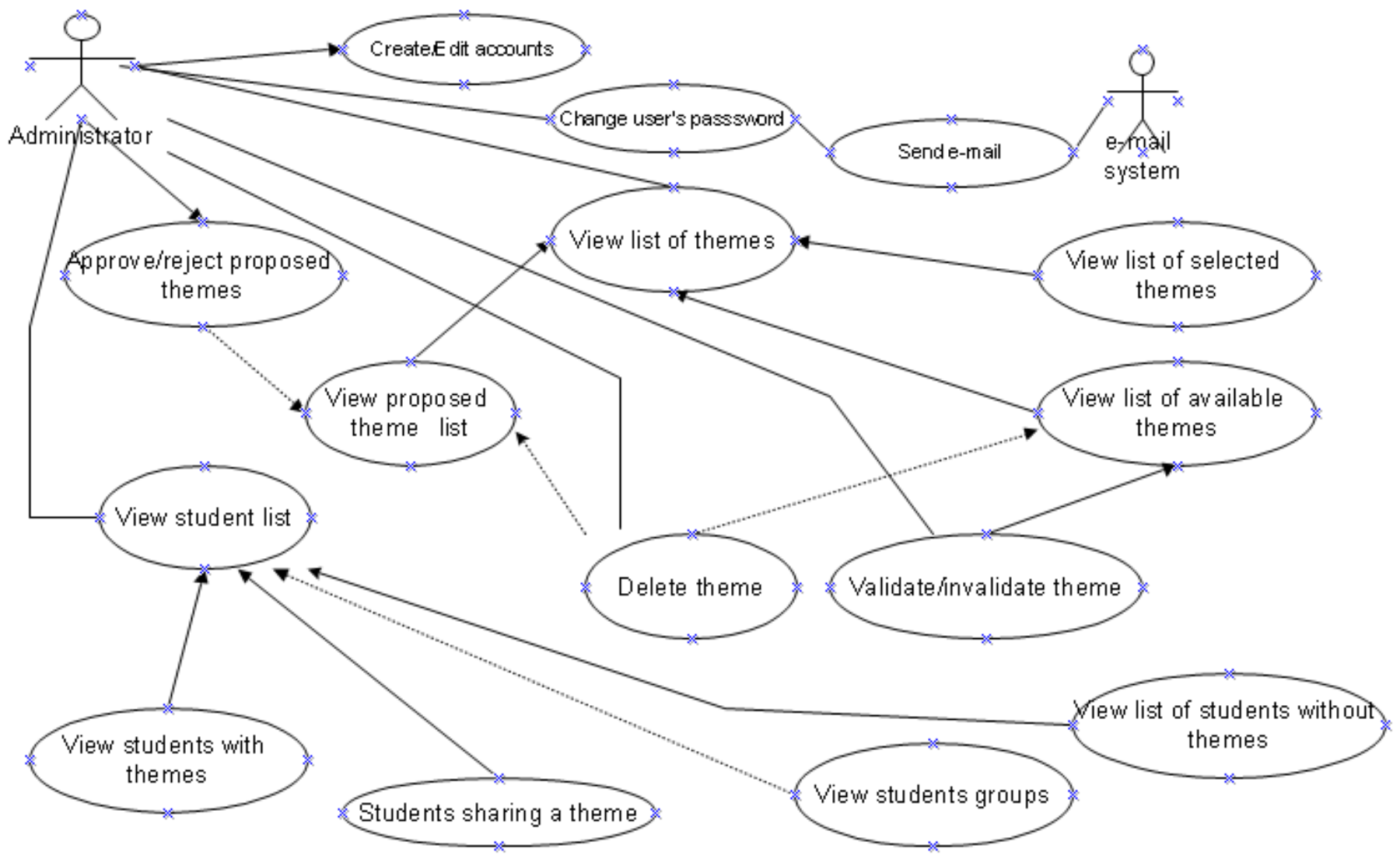

Figure 2. Administrator use case diagram

- **from the users' perspective (figure 3)**
  - the theme list, as well as its files, keywords and associated references, may be viewed by all users;
  - users are entitled to change their passwords.

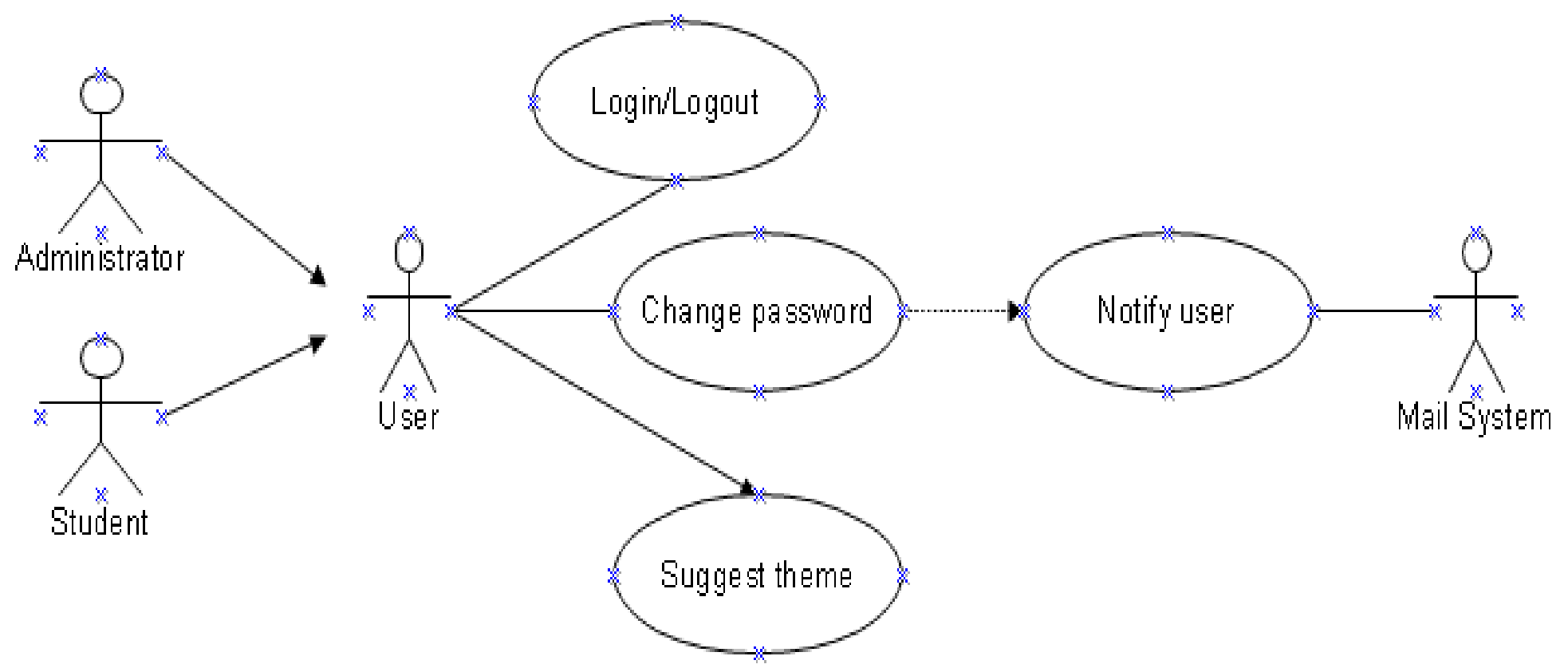

Figure 3. A simplified user case diagram

**Constructing The Entity-Relationship Diagram (ERD)**

An **Entity-Relationship diagram** is a way of representing the structure of a relational database. An entity represents a discrete object. Entities can be thought of (roughly) as nouns. A relationship captures how two or more entities are related to one another.

The ERD, when applied to the specification and high level design of a relational database, performs the following functions:

1. It identifies the tables in a database schema.
2. It identifies the 1 to many relations between tables.
3. It identifies the fields or columns within each table.
4. It designates primary and foreign keys.
5. It may specify data types for each column.

The ERD is a graphical structure. We will be using the MySQL Server to create the tables and relations. Some basic rules for modelling relationships are described in the following paragraphs.

**Entities Must Participate In Relationships**

Entities cannot be modelled if they are not related to any other entity. Otherwise, when the model is transformed into the relational model, there would be no way to navigate to that table. The exception to this rule is a database with a single table.

**Resolve Many-To-Many Relationships**

Many-to-many relationships cannot be used in the data model, because they cannot be represented by the relational model. Therefore, many-to-many relationships must be resolved early in the modelling process. The strategy for resolving many-to-many relationship is to replace the relationship with an *association entity* and then relate the two original entities to the association entity. This strategy is demonstrated below. Figure 6.1 (a) shows the many-to-many relationship:

Students may be **assigned** to many themes.

Each theme may be **assigned** to more than one student.

In addition to the implementation problem, this relationship presents other problems. Suppose we wanted to record information about student assignments such as presentation week. Given the present relationship, these attributes could not be represented in either STUDENT or THEME without repeating information. The first step is to convert the relationship **assigned** to a new entity we will call ASSIGNMENT. Then, the original entities - STUDENT and THEME- are related to this new entity preserving the cardinality and optionality of the original relationships. The solution is shown in Figure 4.

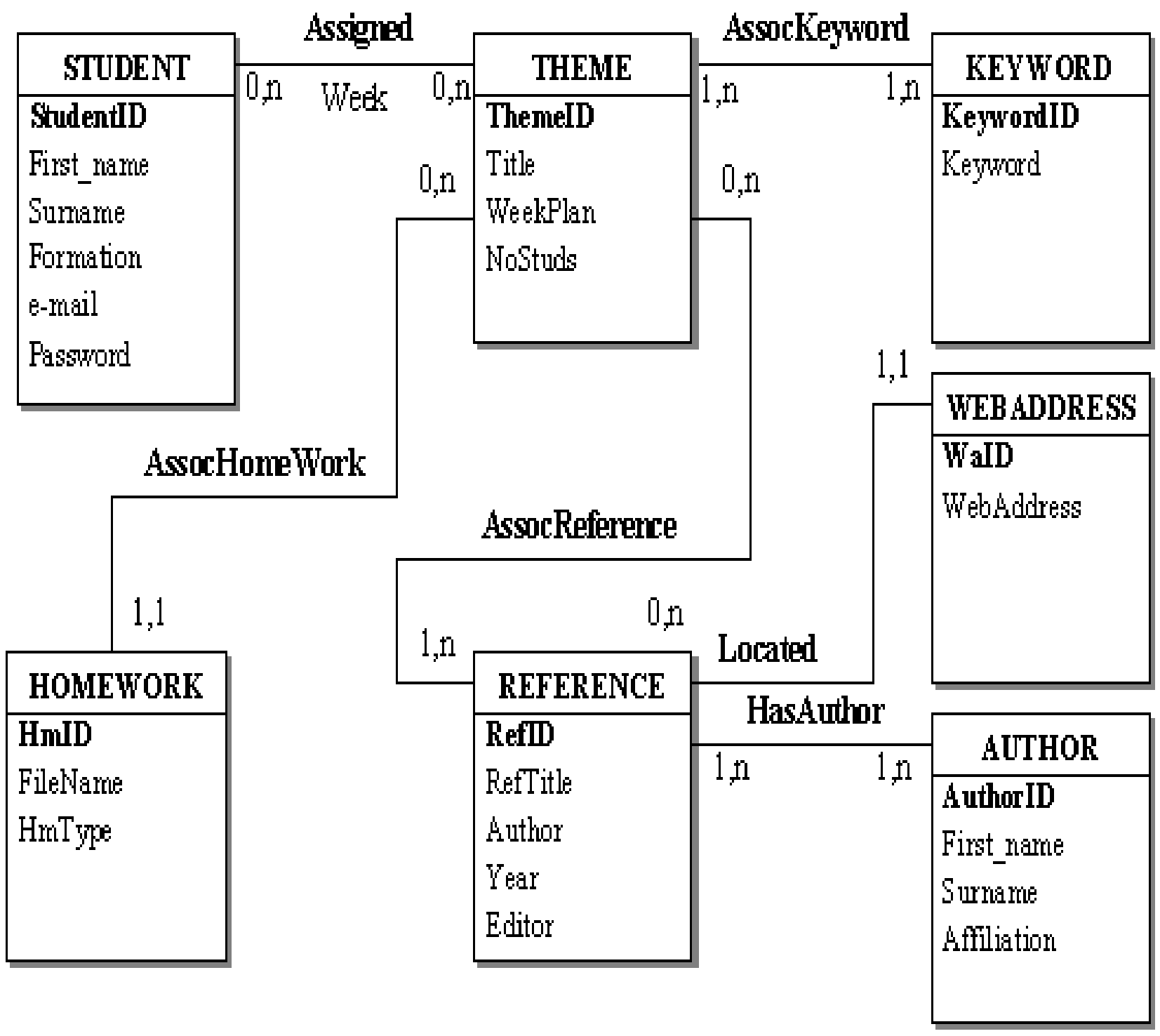

Figure 4. The Entity-Relationship Diagram

Notice that the schema changes the semantics of the original relation to
students **may be assigned** multiple themes
and themes may **be assigned to** more than one student

The implemented database structure is shown in figure 5.

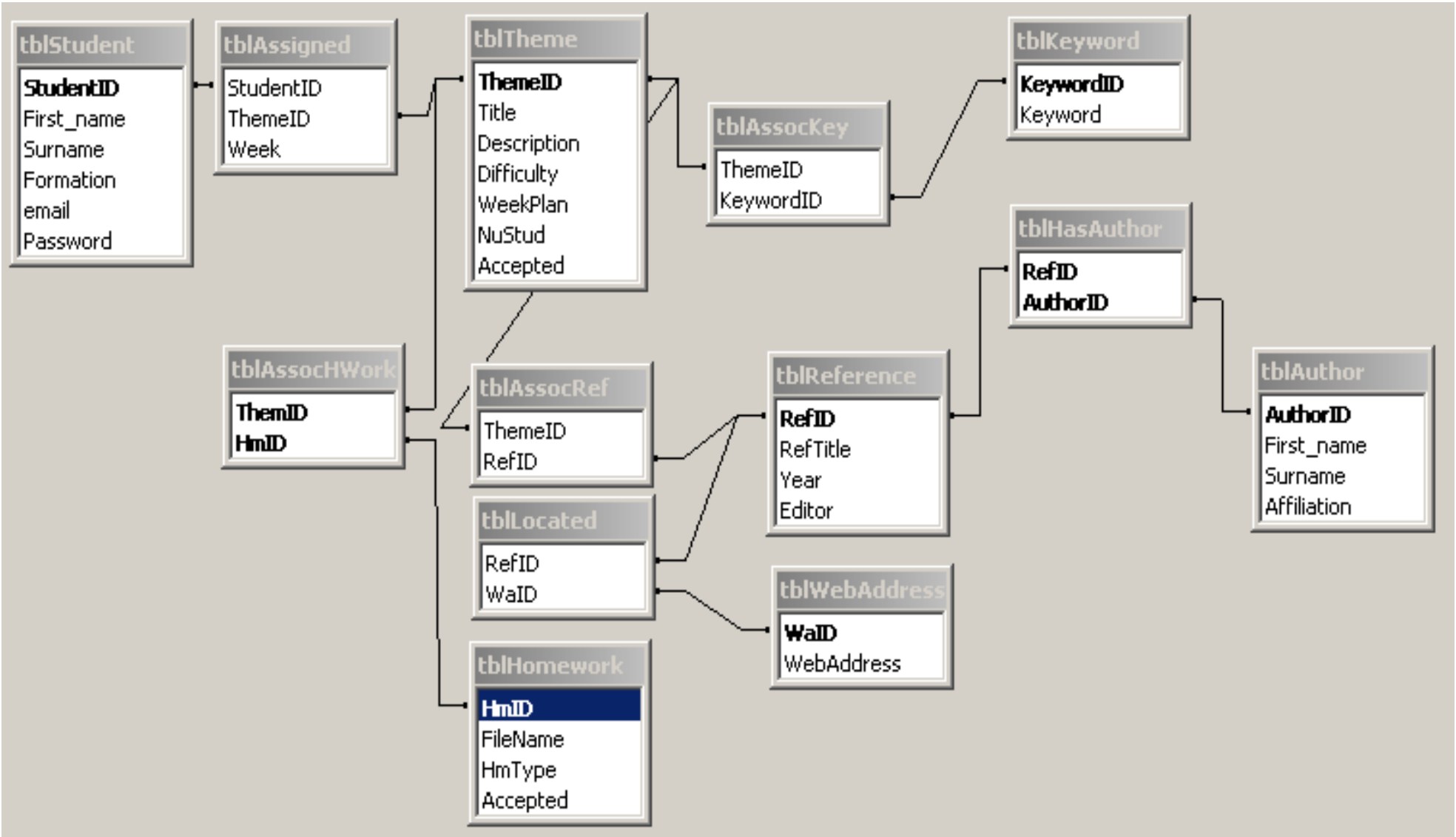

Figure 5. The database structure

**Conclusions**

The present application is being piloted for a relatively reduced number of students (35). The number of themes initially proposed by the teacher did not exceed the corresponding number of students. Nevertheless, within one week the number of proposed and negotiated themes increased to 41.

From the teacher's perspective, the application enhances the teacher-student relationship. From the students' perspective, the application is a complex instructional ensemble designed to transform learning into a round-the-clock activity**.**